\documentclass[final,compress,3p,times]{elsarticle}

\usepackage{lipsum}
\makeatletter
\def\ps@pprintTitle{%
 \let\@oddhead\@empty
 \let\@evenhead\@empty
 \def\@oddfoot{}%
 \let\@evenfoot\@oddfoot}
\makeatother

\usepackage{graphicx}
\usepackage{amssymb}
\usepackage{amsmath}
\usepackage{multicol}

\usepackage{lineno}
\usepackage{xcolor}
\usepackage{hyperref}

\begin{document}

\begin{frontmatter}


\title{Probing Late-Stage Hadronic Interactions at High Baryon Density via $K^{*0}$ Production in the RHIC Beam Energy Scan Program}




\author{
B.~E.~Aboona$^{59}$,
J.~Adam$^{17}$,
G.~Agakishiev$^{32}$,
I.~Aggarwal$^{45}$,
M.~M.~Aggarwal$^{45}$,
Z.~Ahammed$^{66}$,
A.~Aitbayev$^{32}$,
I.~Alekseev$^{3,41}$,
E.~Alpatov$^{41}$,
A.~K.~Alshammri$^{33}$
% ...
}

\address{\rm{(STAR Collaboration)}}
% ...
\address{$^{3}$Alikhanov Institute for Theoretical and Experimental Physics NRC "Kurchatov Institute", Moscow 117218}
% ...
\address{$^{17}$Czech Technical University in Prague, FNSPE, Prague 115 19, Czech Republic}
% ...
\address{$^{32}$Joint Institute for Nuclear Research, Dubna 141 980}
\address{$^{33}$Kent State University, Kent, Ohio 44242}
% ...
\address{$^{41}$National Research Nuclear University MEPhI, Moscow 115409}
% ...
\address{$^{45}$Panjab University, Chandigarh 160014, India}
% ...
\address{$^{59}$Texas A\&M University, College Station, Texas 77843}
% ...
\address{$^{66}$Variable Energy Cyclotron Centre, Kolkata 700064, India}
% ...

\begin{abstract}
A precision measurement of the $K^{*0}$ meson yield is reported in Au+Au collisions at 
$\sqrt{s_{NN}} = 7.7,\; 11.5,\; 14.6,\; 19.6,$ and $27~\mathrm{GeV}$ using the high-statistics 
data sample collected by the STAR experiment during the Beam Energy Scan II (BES-II) 
program at RHIC. The transeverse momentum ($p_{T}$)-integrated yield ratios 
$(K^{*0} + \overline{K^{*0}})/(K^{+} + K^{-})$ in central collisions show a suppression 
relative to peripheral collisions at the $(1.7\text{--}3.6)\,\sigma$ level, while a thermal 
model without final-stage rescattering overpredicts this ratio with a deviation of 
$(6.9\text{--}8.2)\,\sigma$. These results indicate a loss of the measured $K^{*0}$ signal 
in central collisions due to re-scattering of its hadronic decay products in the hadronic phase. The $p_{T}$-integrated yield of charged kaons exhibits an approximate scaling with 
charged-particle multiplicity, independent of collision energy and system size. A similar 
trend is observed for the short-lived $K^{*0}$ resonance, although significant deviations 
emerge at lower energies. At BES energies, the $K^{*0}/K$ ratio shows stronger 
suppression than at the highest RHIC and LHC energies within a given multiplicity bin, 
particularly in central and mid-central collisions. This behavior is consistent with 
changes in the effective hadronic interaction cross section and is supported by transport 
model calculations, which indicate dominant meson--baryon interactions at lower energies 
and meson--meson interactions at higher energies.

\end{abstract}


\end{frontmatter}
\twocolumn

\section*{Introduction}
\label{sec:intro}
The bulk properties of the medium produced in heavy-ion collisions have been comprehensively studied through measurements of light-flavored hadrons ~\cite{STAR:2008med,STAR:2015vvs,STAR:2017ieb,STAR:2017sal,STAR:2019vcp,STAR:2019bjj,Chen:2024aom}.
As the hot and dense medium created in these collisions cools, it transitions to a hadron resonance phase.  During the evolution of heavy-ion collisions and subsequent cooling, the system first reaches a spacetime surface known as the chemical freeze-out, 
where the chemical composition of the medium is fixed,
and inelastic collisions between the constituents cease~\cite{Rafelski:2001hp,Rapp:2000gy,Song:1996ik,Becattini:2005xt,Cleymans:2004bf,Andronic:2005yp,Chen:2024aom}. Later, at the kinetic freeze-out surface, all elastic interactions stop and the momenta of the hadrons are frozen; which are then freely stream to the detectors. Short-lived resonances which decay via strong interactions are excellent probes of this hadronic phase~\cite{Brown:1991kk}. Previously, several resonances such as $\rho^0 (770)$, $K^{*0,*\pm} (892)$, $\Sigma^{* } (1385)$, $\Lambda^{*} (1520)$, and $\phi$ (1020) was studied in $p+p$, $p$+A and A+A collisions at the Relativistic Heavy-Ion Collider (RHIC)~\cite{STAR:2002npn,STAR:2004bgh,STAR:2004rho,STAR:2006vhb,STAR:2008twt,STAR:2010avo,STAR:2023aks,PHENIX:2014kia} and the Large Hadron Collider (LHC)~\cite{ALICE:2012pjb,ALICE:2014jbq,ALICE:2016sak,ALICE:2017ban,ALICE:2019etb,ALICE:2019hyb,ALICE:2019xyr,ALICE:2021ptz,ALICE:2021uyz}. These resonances, with lifetimes ranging 
from 1.2 to 46 fm/$c$~\cite{ParticleDataGroup:2020ssz}, allow a mapping of the hadronic phase of heavy-ion collisions at various time scales~\cite{Brown:1991kk,Bleicher:2002dm,Rafelski:2001hp,Markert:200892,Cabrera:2019,SHAPOVAL2017391}.

 With resonance lifetimes comparable to the duration of the hadronic phase, their decay daughters primarily engage in two types of in-medium interactions: re-scattering and regeneration. If at least one decay daughter interacts elastically with other hadrons in the medium,
the four-momentum information of the decay daughter is altered, compromising the reconstruction of the parent resonance. Concurrently, pseudo-elastic scattering between hadrons in the medium can lead to the regeneration of the resonance state (e.g., $\pi^{-}K^{+} \rightarrow K^{*0} \rightarrow \pi^{-}K^{+}$), resulting in an enhancement of their yield. The strength of these two effects can depend on various factors, such as the time span between chemical and kinetic freeze-out, the lifetime of the resonances, the hadronic interaction cross-section of the decay daughters, and the particle density of the medium. Consequently, the final resonance yield can be modified by these two competing in-medium effects~\cite{Bleicher:2002dm,Knospe:2015nva,Steinheimer:2017,Bleicher:2004}. The relative dominance of one effect over the other can be investigated by examining the ratio of resonance to non-resonance particles with similar quark content, 
as a function of the charged-particle multiplicity of each collision event.

The $K^{*0}$ resonance has a short lifetime of 4.16 fm/$c$ and decays to charged kaons and pions with a branching ratio of 66.66\%. The decay daughters (primarily pions) undergo scattering within the medium, thus reducing the $K^{*0}$ yield. At the same time, the weaker $\pi-K$ interaction cross-section (five times smaller than the $\pi-\pi$ cross-section~\cite{Protopopescu:1973sh,Matison:1974sm,Bleicher:1999xi}) renders the regeneration of $K^{*0}$ from pseudo-elastic scattering negligible. As a result, the $K^{*0}/K$ ratio in central (larger geometric overlap) heavy-ion collisions is lower compared to peripheral (smaller overlap) and small system collisions (e.g., $e+e$, $p+p$, and $p+A$). This observation has been well established by previous experimental studies~\cite{STAR:2002npn,STAR:2008twt,STAR:2010avo,STAR:2023aks,NA61SHINE:2020czr,ALICE:2012pjb,ALICE:2019xyr}. 

The latest results from the STAR Beam Energy Scan (BES) phase-I program are noteworthy, in that the $K^{*0}/K$ ratios at BES energies indicate more suppression compared to the top RHIC and LHC energies within a common multiplicity bin, especially in central and mid-central collisions~\cite{STAR:2023aks}. However, due to large measurement uncertainties, firm conclusion cannot be made. Model calculations that incorporate only meson-meson scattering fail to explain the data at BES energies~\cite{LEROUX2021136284}. Nonetheless, recent work using a transport model approach qualitatively reproduces the trend, that may be driven by changes in the type of QCD medium produced and the hadron interaction cross-sections at top RHIC and BES energies~\cite{Sahoo:2024urqmd}. 

In this Letter, we report precise measurements of $K^{*0}$ mesons at midrapidity $(|y| < 1.0)$ using data from Au+Au collisions at a nucleon-nucleon center-of-mass energy $\sqrt{s_{NN}}$ = 7.7, 11.5, 14.6, 19.6, and 27 GeV, collected by the STAR experiment during 2018-2021 as part of the BES-II program. Throughout this paper, the $K^{*0}(\overline{K^{*0}})$ and $K^{\pm}$ are combined and generically denoted by $K^{*0}$ and $K$, respectively. The organization of this letter is as follows: Section 2 describes the STAR experimental setup along with the applied event and track selection criteria. In Section 3, the data analysis technique, yield extraction procedure, and systematic uncertainties are detailed. Section 4 presents the results from the analysis. Finally, the article concludes with a summary in Section 5.

\section*{Experimental Setup}
\label{sec:expt}
The results presented here are based on data taken with the STAR detector at RHIC for minimum-bias~\cite{STAR:2017sal} 
Au+Au collisions at $\sqrt{s_{NN}}$ from 7.7 to 27 GeV. A detailed description of the STAR sub-detector systems can be found in Ref.~\cite{STAR:2002eio}.
A description of the upgrades for the BES-II program can be found in Ref.~\cite{STARiTPCUpgrade}.  

The Time Projection Chamber (TPC) is the primary tracking device at STAR ~\cite{Anderson:2003ur}.
In 2019, the inner TPC (iTPC) was upgraded to enhance track reconstruction and to extend pseudorapidity coverage. The TPC measures 4.2 meters in length and 4 meters in diameter, covering approximately $\pm$1.5 units of pseudorapidity ($\eta$)
and the full $2\pi$ azimuthal angle. The TPC operates in a nearly constant magnetic field of 0.5 T, oriented in the longitudinal ($z$) direction. Following the upgrade, the TPC can now identify particles with momenta as low as 60 MeV/$c$ (previously 150 MeV/$c$) by utilizing ionization energy loss ($dE/dx$) and momentum information. The Time of Flight (TOF) detector ~\cite{Llope:2003ti,Bonner:2003bv} is composed of rectangular chambers surrounding the TPC, employing Multi-gap Resistive Plate Chamber (MRPC) technology, with an acceptance of $|\eta| < 0.9$ units. The distribution of $dE/dx$ bands for pions and kaons begin to merge in the intermediate $p_T$ range ($p_T\sim 0.6$ GeV/$c$). The TOF detector extends particle identification (PID) capability for charged particles to higher $p_{T}$ ($p_T\sim 1.6$ GeV/$c$).


Collision events are selected by requiring the primary vertex position ($V_{z}$) to be within
$\pm 145$ cm ($\pm$50 cm for $\sqrt{s_{NN}}$ = 27 GeV) with respect to the nominal center of the TPC along the beam direction. To reject background events involving interactions with the beam pipe (radius 3.95 cm), the vertex in the radial direction (defined as $V_{r} = \sqrt{V_{x}^{2} + V_{y}^{2}}$, where $V_{x}$ and $V_{y}$ are the vertex positions along the $x$ and $y$ directions) is required to be within 2 cm of the center of STAR for all energies~\cite{STAR:2008med,STAR:2017sal}. A detailed comparison of tracks recorded in the TPC with those having valid TOF information is performed to remove pile-up events from different bunch crossings, i.e., extra events recorded by the TPC that are not associated with the triggered events.
The minimum-bias triggered~\cite{Llope:2003ti,Adler:2000bd} events, which sample an 
inclusive set of inelastic collisions, are categorized into nine centrality classes based on a ``reference multiplicity" defined by the STAR experiment. This reference multiplicity is calculated from the raw multiplicity of charged particle tracks reconstructed in the TPC across $2\pi$ azimuthal range and within $|\eta| < 0.5$. Centrality reflects the geometrical overlap of the colliding nuclei and is expressed as a percentage of the total inelastic hadronic nucleus–nucleus cross-section, with 0\% corresponding to the most central collisions. The total number of minimum-bias events analyzed for Au+Au collisions at various collision energies, after applying all event selection cuts, are listed in Table ~\ref{tab:datasets}.

\begin{table}[ht]
\begin{center}
    
\begin{tabular}{ |c|c|c|c|c|c| } 
\hline
         $\sqrt{s_{NN}}$ (GeV) & 7.7 & 11.5 & 14.6 & 19.6 & 27 \\
\hline
         No. of Events ($\times 10^{6}$) &   90 & 330 & 394 & 775 & 432 \\
\hline
\end{tabular}
\caption{Total number of minimum-bias events analyzed for Au+Au collisions at various collision energies.}
\label{tab:datasets}
\end{center}
\end{table}

The full coverage of the TPC is utilized by including tracks within $|\eta| < 1.5$ ($|\eta| < 1.0$ for 27 GeV, since it doesn't include iTPC). The distance of closest approach of the selected tracks to the primary vertex must be less than 2 cm to reduce contamination from secondary particles (e.g. weak decay contributions). To avoid short tracks, at least 15 hit points in the TPC are required. To eliminate multiple counting of split tracks, each accepted track must have at least 55\% of the maximum number of possible hit points  in the TPC along its trajectory, that the track could possibly contain~\cite{STAR:2023aks}.

Particle identification is performed using both the TPC and TOF detectors. In the TPC, charged hadrons are identified by their specific energy loss ($dE/dx$), and characterized by an $N\sigma$ variable whose definition is as follows,

\begin{equation}
N\sigma (\pi,K) = \frac{1}{R_{TPC}} \mathrm{log} \frac{(dE/dx)_{\mathrm{ measured}}}{\langle dE/dx \rangle_{\mathrm{theory}}}, \hspace{-.5em}
\label{eqn-nsigma}
\end{equation}
where, $(dE/dx)_{\mathrm{ measured}}$ is the specific energy loss measured in the TPC for a given track, $\langle dE/dx \rangle_{\mathrm{theory}}$ is the expected mean energy loss calculated from the Bethe–Bloch parametrization for a given particle species and momentum~\cite{Bichsel:2006cs}. and $R_{TPC}$ is the $dE/dx$ resolution of the TPC detector. 
With the inclusion of the iTPC during the BES-II program, the no. of pad rows in the inner TPC sector increased, leading to a higher number of ionization samples per track. As a result, the path-length dependent dE/dx resolution improved from about 7.5\% in BES-I to approximately 6.5\% in BES-II, enhancing particle identification capability at low transverse momentum~\cite{STARiTPCUpgrade}.
Pions and kaons are selected with a TPC $|N\sigma| < 2$ requirement. The time-of-flight information for each track is obtained using the TOF detector. This allows the determination of the particle mass-squared ($m^{2}$), which is then used for particle identification by applying suitable mass-squared cuts.
In this analysis, we used both the TPC and TOF detectors for particle identification.
 When TOF information is available, particles are identified 
 using the squared mass ($m^{2}$) within the ranges $-0.2 < m_{\pi}^{2} < 0.15$ and $0.16 < m_{K}^{2} < 0.36$ (GeV/$c^{2}$)$^{2}$, respectively~\cite{STAR:2023aks}.


\begin{figure*}[!ht]
\begin{center}
\includegraphics[scale=0.65]{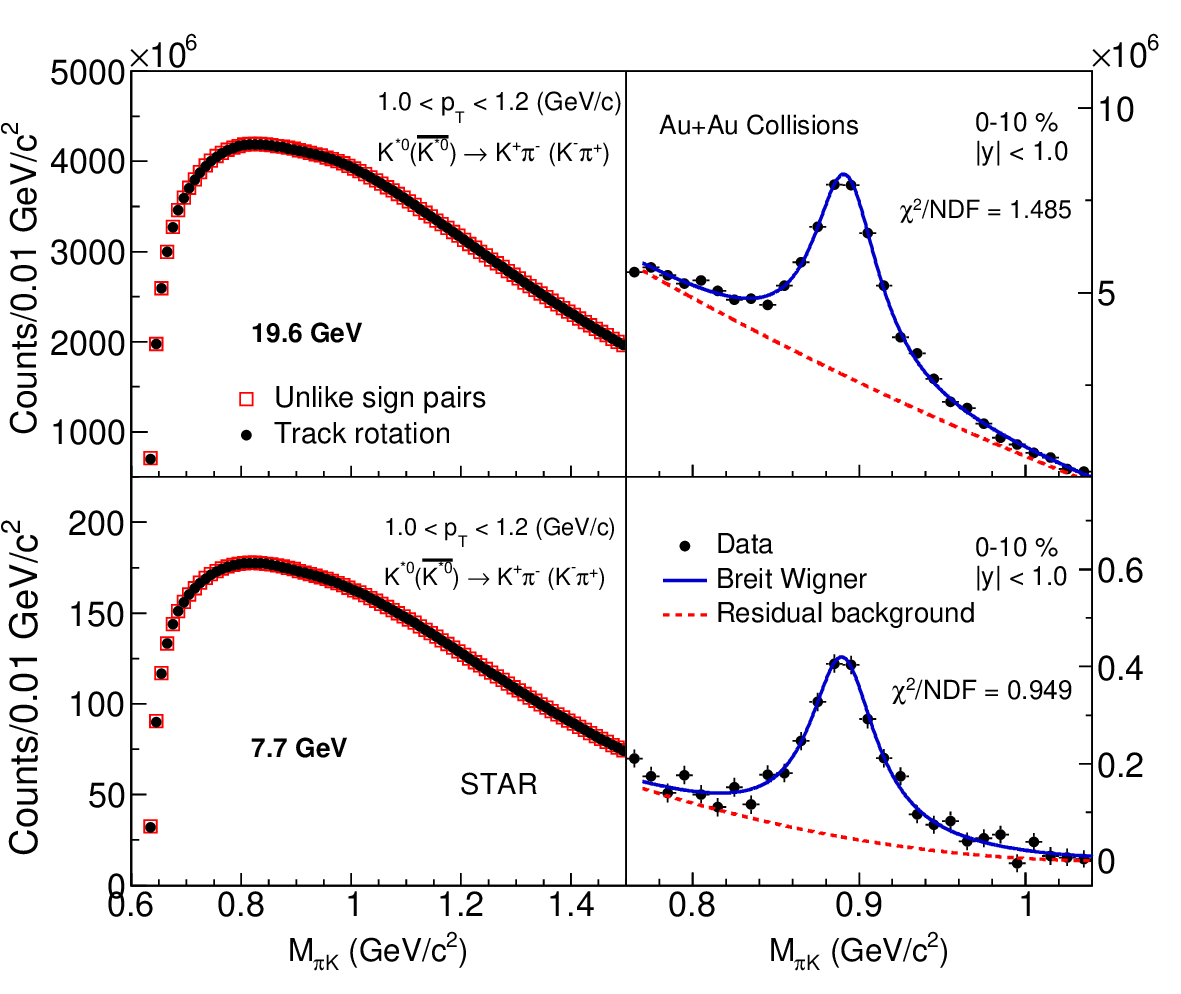}

\caption{(Left panels) $K^{*0}$ signal from unlike-sign pairs (red markers) and background estimated from the track rotation method (black markers). (Right panels) Invariant mass distribution of $K\pi$ pairs after subtraction of the estimated background using the track rotation method. The blue line denotes the Breit Wigner fit and the dashed red line represents the residual background.}
\label{fig:signal}
\end{center}
\end{figure*}

\section*{Data Analysis}
\label{sec:ana_detail}
\subsection*{Yield Extraction}
\label{subsec:yield_extract}
In this analysis, the $K^{*0}$ meson is reconstructed from its hadronic decay channel, $K^{*0}(\overline{K^{*0}}) \rightarrow K^{\pm}\pi^{\mp}$.
The invariant mass distribution of oppositely charged kaons and pions is constructed using pairs from the same event (unlike sign pairs) to reconstruct the resonance signal. This distribution contains the $K^{*0}$ signal peak along with a large, uncorrelated combinatorial background, that is estimated using the track-rotation method. In this technique the momentum vector of one of the daughter tracks (in this case, the pion) is rotated by 180 degrees in the transverse plane to remove the correlation between daughter pairs originating from the same mother particle. 
After subtracting the combinatorial background, the signal peak is observed atop a residual background. This residual background arises from other sources such as correlated background pairs from jets, decays of other resonances, or misidentified daughter tracks. The residual background exhibits a smooth dependence on mass, and its shape is well described by a second-order polynomial~\cite{ALICE:2014jbq,ALICE:2017ban}.

In Fig. ~\ref{fig:signal}, the left panels show the invariant mass distributions of $\pi^{\pm}K^{\mp}$ pairs from the same events and the background estimated from the track rotation method in the transverse momentum interval $1.0 < p_{T}^{K^{*0}} < 1.2$ GeV/$c$ for 0-10\% most central Au+Au collisions at $\sqrt{s_{NN}} =$ 7.7 and 19.6 GeV. The $K^{*0}$ mass peak, obtained after subtracting the combinatorial background, is shown in the two right-hand panels of Fig.~\ref{fig:signal}. The statistical significance of the $K^{*0}$ peak presented here has been improved by a factor of four compared to previously published results using BES-I data~\cite{STAR:2023aks}. 

The signal peak is fitted with a Breit-Wigner (BW) function and the residual background with a second-order polynomial function. The fit is performed in the range $0.77 < M_{\pi K} < 1.04$ GeV/$c^{2}$. During the fitting procedure the width of the $K^{*0}$ meson is fixed to its vacuum values ($\Gamma [K^{*0}] = 47.4 \pm 0.6$ MeV/$c^{2}$~\cite{ParticleDataGroup:2020ssz}). The sensitivity to the choice of the fitting range, the shape of the background function, and the width parameters are studied by varying the default settings as described later in this section. Raw yields of $K^{*0}$ in different $p_{T}$ intervals and event centrality classes are obtained from the integral of the signal peak after removing the residual background as described in Refs. ~\cite{STAR:2010avo,STAR:2023aks}. 

\begin{figure*}[!ht]
\begin{center}
\includegraphics[scale=0.73]{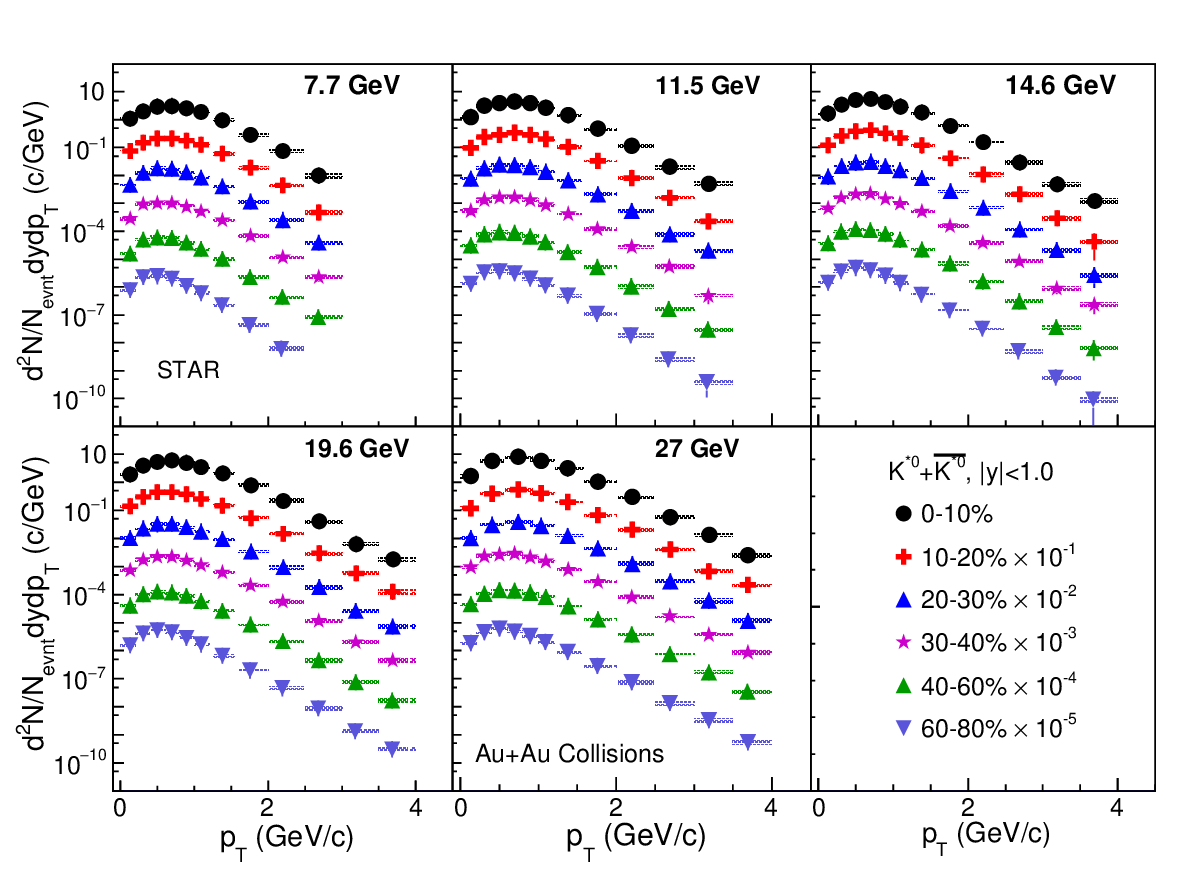}

\caption{$K^{*0}$ $p_{T}$ spectra at mid rapidity ($|y| < 1.0$) in various centralities in Au+Au collisions at $\sqrt{s_{NN}}$ = 7.7, 11.5, 14.6, 19.6 and 27 GeV. The data-points are placed to the mean position in each $p_T$ bin. The statistical and systematic uncertainties are shown as bars and boxes, respectively.}
\label{fig:spectra}
\end{center}
\end{figure*}


\subsection*{Efficiency $\times$ Acceptance corrections}
The primary correction to the raw spectra compensates for losses due to detector acceptance and the efficiency of reconstructing particle tracks.
These are evaluated using the STAR embedding method~\cite{STAR:2008med,STAR:2017sal}. Monte Carlo simulated $K^{*0}$ mesons, distributed uniformly in rapidity ($|y| < 1.5$), transverse momentum ($0 < p_{T} < 6$ GeV/$c$), and azimuthal angle ($0 < \phi < 2\pi$), are embedded into real event raw data. The number of embedded $K^{*0}$ mesons per event was approximately 5\% of the measured charged particle multiplicity for that event. 

The embedded particles are assumed to originate from the measured primary vertex for each event. Successive propagation of $K^{*0}$ meson through STAR, their decays, and the propagation of the daughter particles are simulated using the GEANT3 package~\cite{STAR-GEANT}. 
The simulated electronic signals are combined with those from the real event and processed using the standard STAR reconstruction algorithm. Finally, the reconstruction efficiency $\times$ acceptance is determined by dividing the number of reconstructed $K^{*0}$ mesons, after passing through the detector simulations using the same track/event selection criteria, as used in real data analysis, by the number of simulated MC $K^{*0}$ within the identical rapidity interval.
The tracking efficiency depends on the final-state particle multiplicity in the event, which can vary from a few tracks in peripheral collisions to about a thousand tracks in most-central collisions. Consequently, efficiencies increase by a small amount toward peripheral collisions, as compared to central collisions, owing to decreased track multiplicity.

The choice of TPC $N \sigma$ and TOF mass-squared cuts on candidate $K^{*0}$ daughters may also result in the loss of daughter tracks. Hence the $p_{T}$ spectra are further corrected by a $p_{T}$ dependent PID efficiency. The PID efficiency is calculated in a data driven way~\cite{STAR:2017sal}. The combined TPC($dE/dx$) and TOF PID efficiency, for reconstructed tracks, is 91\%.



\begin{figure*}[!ht]
\begin{center}
\includegraphics[scale=0.7]{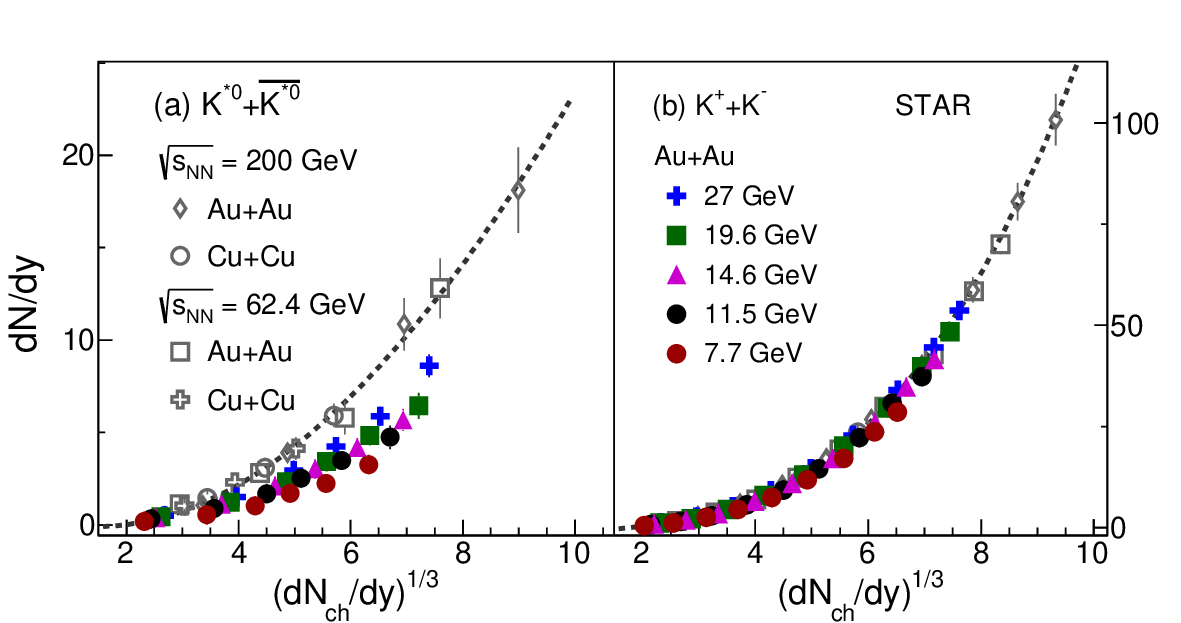}

\caption{The $p_{T}$-integrated yield of $K^{*0}$ (left panel) and charged kaons (right panel) at mid-rapidity as a function of $(dN_{ch}/dy)^{1/3}$ for various collision systems and collision energies~\cite{STAR:2008med,STAR:2017sal,STAR:2019vcp,BRAHMS:2016cucu}.For the $K^{*0}$ measurement, a rapidity window of $|y|<1.0$ is used at BES energies and $|y|<0.5$ at $62.4$ and $200$~GeV. In contrast, kaon measurements are done within $|y|<0.1$ for all collision systems and energies.
 The error bars shown here are the quadrature sum of statistical and systematic uncertainties. The dotted line shown on both the panels are the fit to 200 GeV data with a polynomial function. 
}
\label{fig:dndy_dN13}
\end{center}
\end{figure*}

\label{subsec:effi}
\subsection*{Systematic uncertainties}
\label{subsec:syst}
Systematic uncertainties for the $K^{*0}$ yields are evaluated for each centrality and $p_{T}$ bin. The sources of these uncertainties are discussed below.

The systematic uncertainties associated with signal extraction are estimated by varying the invariant mass fitting range and the form of the residual background function (\textit{i.e.} first order versus second order polynomials). As a cross check, the combinatorial background is also estimated using the like-sign method. In this approach tracks with the same charge from the same event are paired to estimate the uncorrelated background contribution. The invariant-mass distribution for the uncorrelated background is calculated as the geometric mean, $2\sqrt{N^{++} \times N^{--}}$, where $N^{++}$ and $N^{--}$ represent the numbers of positive-positive and negative-negative pairs in each invariant mass interval, respectively. The difference in the yields, calculated from the two background estimation methods was considered in the systematic uncertainty estimates. The systematic uncertainty in the yield calculation is determined using two different methods: histogram integration and fitted BW function integration. Variations in the yields, calculated by keeping the width fixed to the PDG value, or allowing the width to be a free parameter, are also included in the systematics.
In addition, the criteria for event and track quality cuts, as well as for particle identification selections, are varied to evaluate the associated systematic uncertainties. The uncertainty arising from track reconstruction efficiency is estimated to be 4.8\% for the $K^{*0}$ meson daughter pairs.
The final systematic uncertainty is the quadratic sum of the systematic uncertainties from each of the aforementioned 
sources. The typical, average systematic uncertainties in the $K^{*0}$ yields from these sources are listed in table~\ref{tab:syst}.

\begin{table}[ht]
\begin{center}
    
\begin{tabular}{ |c|c| } 
\hline
         Sources of systematic uncertainty  & $K^{*0}$ Yield (\%)\\
\hline
         Fitting region  & 1 \\
\hline
         Residual background shape & 2-3\\
\hline
         Yield extraction method & 3-6\\
\hline
         Track selection  & 1\\
\hline
         Width fix/free  & 3-5\\
\hline
         Particle identification  & 1-2\\
\hline
         Global tracking efficiency  & 4.8\\
\hline
         Total  & 7-10\\
\hline
\end{tabular}
\caption{The sources of systematic uncertainties for $K^{*0}$ yield in Au+Au collisions at $\sqrt{s_{NN}}$ = 7.7 to 27~GeV. The total systematic uncertainty is the sum of the uncertainty from each source in quadrature.}
\label{tab:syst}
\end{center}
\end{table}

\section*{Results}
\label{sec:results}
\subsection*{Transverse Momentum Spectra}
\label{subsec:spectra}

The measured $K^{*0}$ meson $p_{T}$ distributions for $|y| < 1.0$, normalized to the number of events and corrected for detector acceptance and reconstruction efficiency ($\epsilon_{acc \times rec}$), PID efficiency ($\epsilon_{PID}$), and branching ratio (B.R.) of the decay channel, are shown in Fig.~\ref{fig:spectra}. 
The typical value of the reconstruction efficiency 
ranges from 20\% (low $p_T$) to 70\% (high $p_T$). The results for Au-Au collisions are presented for six different centrality classes (0-10\%, 10-20\%, 20-30\%, 30-40\%, 40-60\% and 60-80\%). The lowest $p_{T}$ bin in the measured $p_{T}$ spectra is $0-0.2$ GeV/$c$, while the full $p_T$ range extends to 4 GeV/$c$, expanding the BES-I range of $0.4-3.0$ GeV/$c$. The increased $p_T$ range is made possible by the iTPC upgrade and the larger number of events collected during the BES-II program.


\subsection*{$p_{T}$ integrated yield ($dN/dy$)}
\label{subsec:dndy}
The $p_{T}$-integrated particle yield ($dN/dy$) is extracted for each
centrality interval from the measured $p_{T}$ spectrum of $K^{*0}$ mesons. Since the lowest $p_{T}$ bin in the measured $p_{T}$ spectra is $0-0.2$ GeV/$c$, no low-$p_{T}$ extrapolation is required to extract $dN/dy$, however the extrapolation for higher $p_T$ is performed. The extrapolated fraction of the yield is negligible ($< 1\%$) for $p_{T} > 4$ GeV/c.

The Hanbury-Brown Twiss radii have been observed to increase linearly with the  cube root of the charged-particle multiplicity, $(dN_{ch}/dy)^{1/3}$ and we therefore assume that $(dN_{ch}/dy)^{1/3}$, measured at mid-rapidity, can serve as a proxy for the system size~\cite{Hbt:2005}. Here, $(dN_{ch}/dy)$ is the sum of the mid-rapidity $\pi^{\pm}, K^{\pm},\text{and}~p(\Bar{p})$ yields~\cite{STAR:2017sal,STAR:2019vcp}. Fig.~\ref{fig:dndy_dN13} shows the variation of $K^{*0}$ and charged kaons
as a function of $(dN_{ch}/dy)^{1/3}$. The charged kaon yields are taken from the published BES-I measurements~\cite{STAR:2008med,STAR:2017sal,STAR:2019vcp}. The results are compared with various collision systems and beam energies available from RHIC~\cite{STAR:2008med,STAR:2017sal,STAR:2019vcp,STAR:2010avo,STAR:2023aks,BRAHMS:2016cucu}. It should be noted that the collision-energy dependence of the $\overline{K^{*0}}/K^{*0}$ ratio behaves similarly to that of the $K^-/K^+$ ratio at BES energies. Since $K^{*0}$ and K have similar quark content (similar masses for the $u$ and $d$ quark isospin doublet), one would expect very similar strong interaction mechanisms to be involved in their production.
It can be seen that the $dN/dy$ of both $K^{*0}$ and $K$ increases with $(dN_{ch}/dy)^{1/3}$ across all collision systems and beam energies. The charged kaon $dN/dy$ appears to show an approximate scaling with $(dN_{ch}/dy)^{1/3}$, largely independent of the collision system and energy, whereas the $K^{*0}$ yield exhibits a more pronounced violation of this scaling. At higher collision energies (open markers in Fig.~\ref{fig:dndy_dN13}a), the $K^{*0}$ yield appears to show a monotonically increasing dependence on $(dN_{ch}/dy)^{1/3}$ across all systems and energies, similar to the charged-kaon yield. However, towards lower collision energies, a gradual breaking of the scaling behavior is observed for both particles. Quantitatively, the deviation from the scaling for charged kaons is in the range of 1--4$\sigma$, while for $K^{*0}$ it is significantly larger, about 4--10$\sigma$ at the lower BES energies (7.7--27 GeV), i.e., approximately 2--4 times larger than that observed for charged kaons. This enhanced deviation of the $K^{*0}$ yield is likely related to increased loss due to re-scattering in the hadronic phase at lower collision energies.


\begin{figure*}[!ht]
\begin{center}
\includegraphics[scale=0.5]{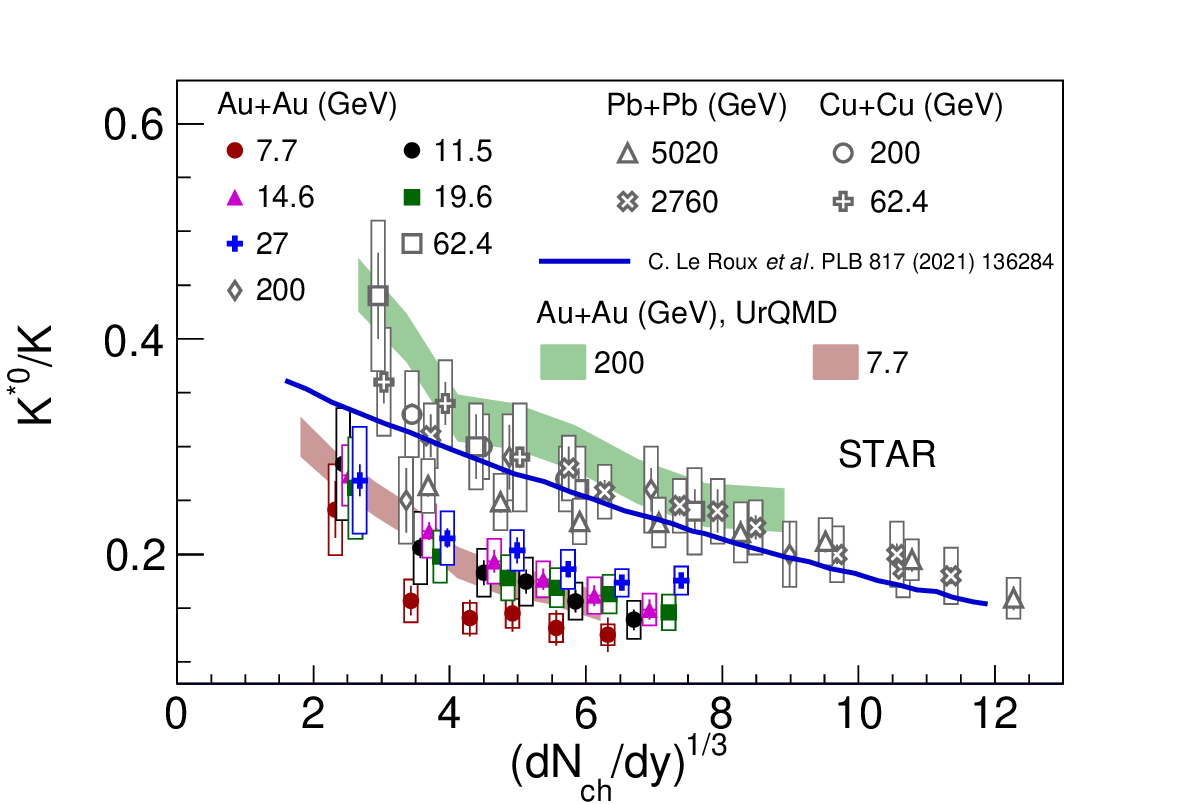}

\caption{$K^{*0}/K$ ratio as a function of $(dN_{ch}/dy)^{1/3}$ in Au+Au collisions. The measurements are compared with previous experimental results. The data points with closed symbols are generated from this analysis. The model calculation (blue solid line) is from ~\cite{LEROUX2021136284}. UrQMD results for $\sqrt{s_{NN}}$ = 7.7 and 200 GeV Au+Au collisions are shown in colored bands and taken from ~\cite{Sahoo:2024urqmd}.  The $K^{*0}(\overline{K^{*0}})$ and $K^{\pm}$ are combined and denoted by $K^{*0}$ and $K$, respectively, for BES energies.}
\label{fig:ratio_dN13}
\end{center}
\end{figure*}

\begin{figure*}[!ht]
\begin{center}
\includegraphics[scale=0.7]{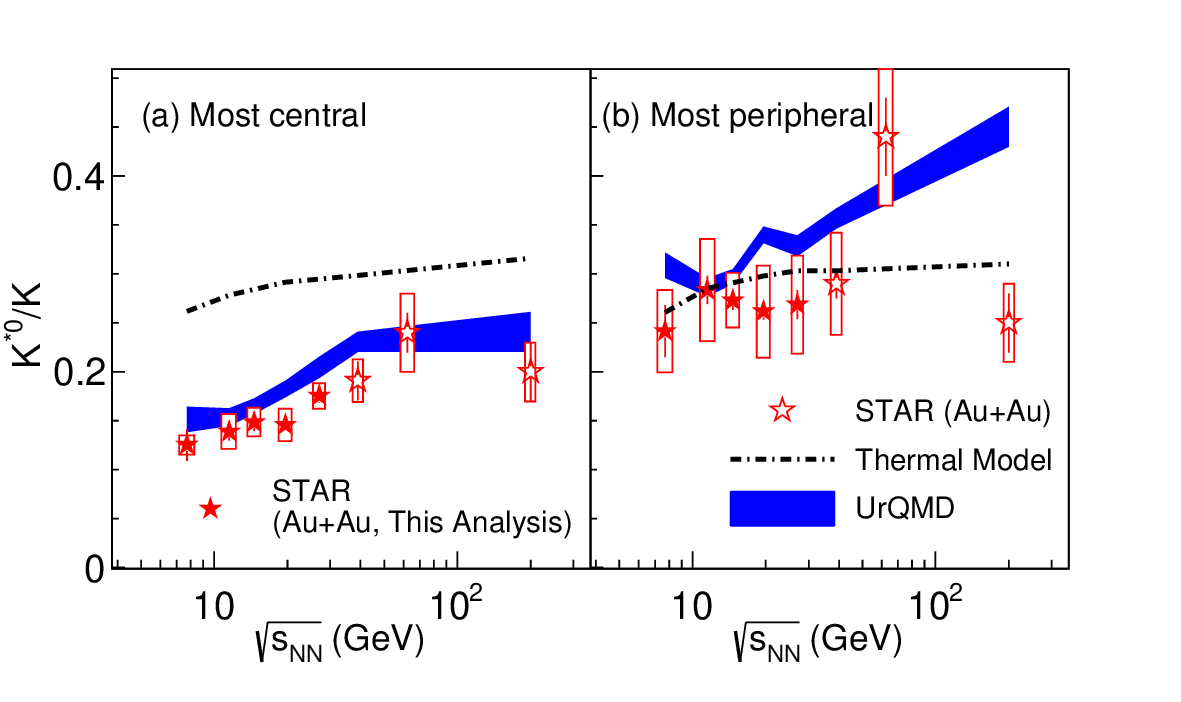}

\caption{$K^{*0}/K$ ratios as a function of $\sqrt{s_{NN}}$ in Au+Au collisions in most-central (left panel) and most-peripheral (right panel) collisions. The measurements are compared with a thermal model~\cite{Sahoo:2023thermal} (dashed line) and the UrQMD model~\cite{Sahoo:2024urqmd} (blue band). The central collisions correspond to 0-10\% centrality (0-20\% for 62.4 GeV) and the peripheral collisions correspond to 60-80\% centrality for all RHIC energies. The $K^{*0}(\overline{K^{*0}})$ and $K^{\pm}$ are combined and denoted by $K^{*0}$ and $K$, respectively, for BES energies.}
\label{fig:ratio_modelcomp}
\end{center}
\end{figure*}

\subsection*{$K^{*0}/K$ ratios}

The lifetime of the $K^{*0}$ meson is comparable to that of the fireball created in relativistic heavy-ion collisions, making it highly susceptible to final-stage effects such as re-scattering and regeneration in the late stage of the hadronic medium. Therefore, the multiplicity dependent yield ratio of $K^{*0}$ mesons to long-lived non-resonant particles, with similar quark content (e.g., kaons), can help elucidate the interplay between these in-medium dynamics. If re-scattering dominates regeneration, the former will impede the reconstruction of short-lived resonances, leading to a reduced $K^{*0}/K$ ratio and vice-versa. This effect is expected to be more pronounced in central collisions, where the hadronic phase can (naively) be assumed to be longer than in peripheral collisions~\cite{STAR:2023aks,STAR:2010avo,ALICE:2012pjb, ALICE:2014jbq}.

Figure~\ref{fig:ratio_dN13} shows, the $K^{*0}/K$ [$=(K^{*0}+\overline{K^{*0}})/(K^{+}+K^{-})$] ratios as a function of $(dN_{ch}/dy)^{1/3}$ for Au+Au collisions at BES energies. 
The charged kaon yields from Refs.~\cite{STAR:2017sal,STAR:2019vcp} are taken within the rapidity window $|y|<$ 0.1. UrQMD studies indicate that using kaon yields in a wider window of $|y|<$ 1.0 would increase the $K^{*0}/K$ ratio by approximately 14\% at 7.7 GeV and by about 4\% at 27 GeV, compared to the reported values.
The $K^{*0}/K$ ratios are compared with previous results from STAR and ALICE~\cite{STAR:2010avo, ALICE:2014jbq, ALICE:2021ptz}.  The $K^{*0}/K$ ratios in central collisions are found to be smaller than those in peripheral collisions, indicating that re-scattering dominates over regeneration in central heavy-ion collisions. This observation agrees with earlier experimental measurements.
The suppression of the $K^{*0}/K$ ratio from peripheral to central collisions at BES-II is similar to that of BES-I~\cite{STAR:2023aks}. In the earlier measurement, the suppression was seen at the $\sim 1.5\sigma$ level, whereas the current measurement observes it with higher precision ($1.7$–$3.6\sigma$). Here the significance is obtained by taking the ratio of the difference between the $K^{*0}/K$ ratio in peripheral to central collisions to the total uncertainties added in quadrature.
This is the first time light-flavored resonance yield suppression of order $3\sigma$ has been observed at RHIC energies.

In Fig. ~\ref{fig:ratio_dN13}, an approximate scaling in the $K^{*0}/K$ ratio at top RHIC and LHC energies (62.4-5020 GeV) can be observed, regardless of the collision system.
However, as the collision energy decreases, the $K^{*0}/K$ ratio is significantly lower compared to that seen at higher energies. 
This deviation in the $K^{*0}/K$ ratio suggests that, even at similar multiplicities, the effects of re-scattering differ at high and low collision energies. Earlier phenomenological work, which only considered the interaction of $K^{*0}$ and $K$ mesons with light mesons in the hadronic medium, describes the multiplicity dependence of the $K^{*0}/K$ ratio at 200 GeV and above. This result indicates that re-scattering is primarily dominated by meson-meson interactions in the high energy regime~\cite{LEROUX2021136284}. The results are also compared with UrQMD model calculations from Ref.~\cite{Sahoo:2024urqmd}. The UrQMD model incorporates a large set of baryon and meson resonances (and their antiparticles and isospin states), allowing for baryon–baryon, meson–baryon, and meson–meson interactions within the hadronic phase~\cite{Bass:1998ca,Bleicher:1999xi}. This transport-model study qualitatively reproduces the observed energy dependence of the $K^{*0}/K$ ratio at BES energies~\cite{Sahoo:2024urqmd}, suggesting that the enhanced baryon-to-meson ratio of yields at lower energies increases the probability of meson–baryon interactions. Additional theoretical insight, including other potential physics mechanisms, is required to better interpret these results.


In Fig.~\ref{fig:ratio_modelcomp}, the $K^{*0}/K$ ratio is shown as a function of collision energy, with comparisons made to both thermal and UrQMD models. The model results are from Ref.~\cite{Sahoo:2023thermal,Sahoo:2024urqmd}. The thermal model, which does not account for the rescattering effects during the hadronic stage,
overestimates the $K^{*0}/K$ ratio at central collisions by $6.9-8.2~\sigma$,
while it remains consistent with the ratios in peripheral collisions. In contrast, the UrQMD model, which includes in-medium particle interactions, agrees better with the data for central collisions, reflecting the presence of strong re-scattering effects. 

\subsection*{Comparison with Blast-wave model:}

\begin{figure*}[!ht]
\begin{center}
\includegraphics[scale=0.73]{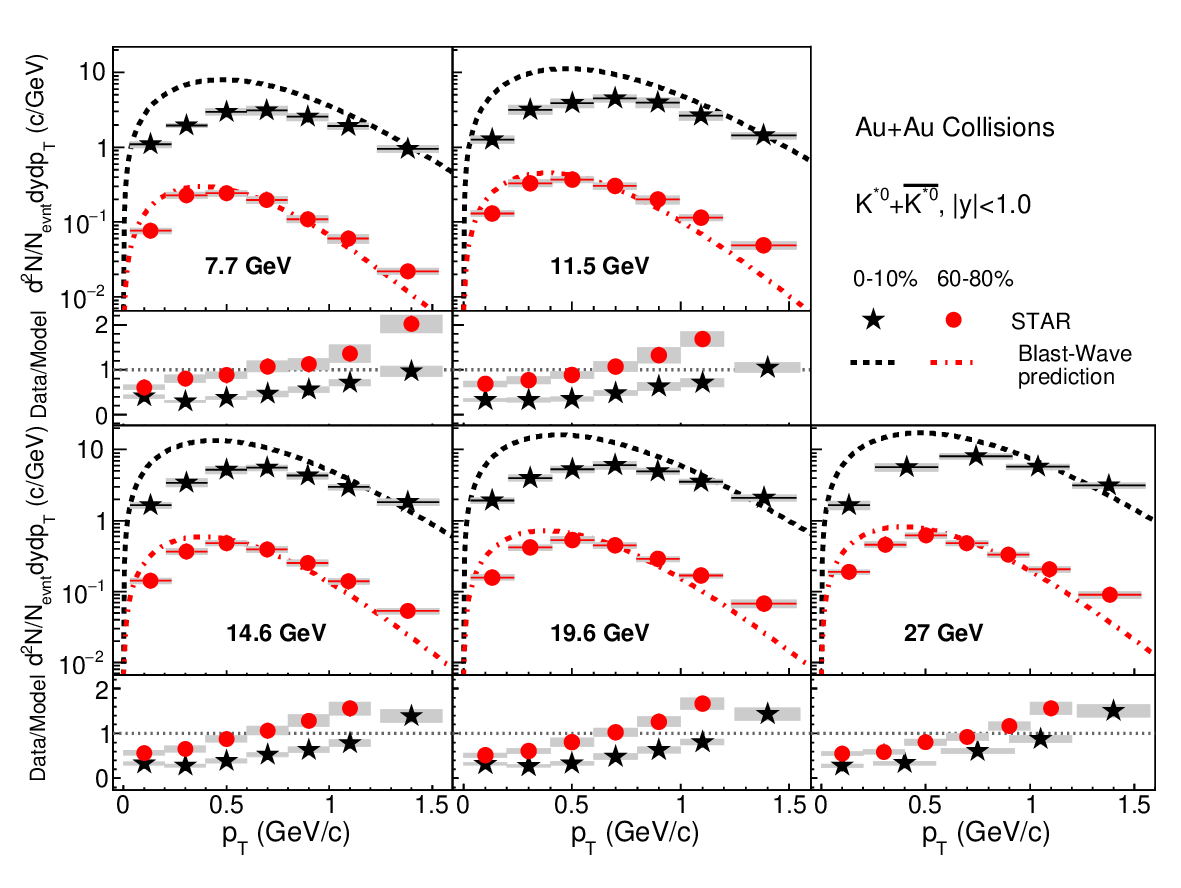}

\caption{The $p_{T}$ distributions of the $K^{*0}$ meson in 0--10\% (black markers) and 60--80\% (red markers) central Au+Au collisions at $\sqrt{s_{NN}} = 7.7$, 11.5, 14.6, 19.6, and 27 GeV, along with the expected $p_{T}$ distributions from the Boltzmann--Gibbs blast-wave function (dotted lines). The blast-wave model is parameterized using parameters obtained from simultaneous fits to the $p_{T}$ distributions of $\pi^{\pm}$, $K^{\pm}$, and $p(\bar{p})$ at respective collision energies. The statistical and systematic uncertainties are represented by bars and shaded boxes, respectively.}
\label{fig:bw_spectra}
\end{center}
\end{figure*}

In order to investigate the $p_{T}$ dependence of the $K^{*0}$ yield suppression, the Boltzmann-Gibbs blast-wave model that does not include resonance decay or rescattering effects is used. This hydrodynamics-inspired model assumes that all particles are emitted from a common kinetic freeze-out surface characterized by a kinetic freeze-out temperature $T_{\mathrm{kin}}$ and move with a common surface expansion velocity $\beta$, often referred to as the transverse radial flow, with a velocity profile exponent $n$. For this study, the freeze-out parameters for different collision energies are taken from Ref.~\cite{STAR:2017sal,STAR:2019vcp}, where the parameters are obtained from simultaneous fits to the $p_{T}$ distributions of $\pi^{\pm}$, $K^{\pm}$, and $p(\bar{p})$ at BES energies. Using these parameters, the expected $p_{T}$ distribution of the $K^{*0}$ meson is calculated. The distributions are normalized such that their integrals are equal to the measured yield of charged kaons times the $K^{*0}/K$ ratio obtained from the thermal model~\cite{Sahoo:2023thermal} at the corresponding collision energies. The resulting $K^{*0}$ $p_{T}$ spectra, without rescattering effects, are obtained from the blast-wave model for central (0--10\%) and peripheral (60--80\%) collisions and are shown as dotted lines in Fig.~\ref{fig:bw_spectra}, together with the $p_{T}$-differential ratio of the measured data to the model prediction. A similar procedure has also been performed in previous studies at LHC energies~\cite{ALICE:2014jbq,ALICE:2019xyr}.

From Fig.~\ref{fig:bw_spectra}, a momentum-dependent depletion of the resonance yield is observed, particularly at $p_{T} < 1.5$ GeV/$c$, similar to that predicted by the UrQMD model~\cite{Bleicher:2002dm,Bass:1998ca,Bleicher:1999xi}. The data-to-model ratio shows a strong deviation from unity, indicating a suppression up to about 70\% for central collisions, whereas relatively small suppression is observed for peripheral collisions. The suppression of the $K^{*0}$ yield with respect to the blast-wave model expectation in central collisions, relative to peripheral collisions, is consistent with scenerio of resonance yield suppression due to the dominance of rescattering effects within the hadronic phase.

\section*{Summary} 
\label{sec:sum}
The $p_{T}$ distributions of $K^{*0}$ mesons are measured at mid rapidity ($|y| < 1.0$) in Au+Au collisions at $\sqrt{s_{NN}}$ = 7.7 to 27 GeV, using the high statistics data collected during the STAR BES-II program. 

At the lower BES-II energies, the $K^{*0}$ meson yields deviate from the multiplicity scaling, suggesting the presence of additional hadronic dynamics, such as enhanced baryon-meson re-scattering, which can reduce the observed resonance yield. In contrast, the charged kaon ($K^{\pm}$) $p_{T}$-integrated yield follows a multiplicity-based scaling across the RHIC collision energy range from 7.7 to 200 GeV.

Furthermore, the ratio of $K^{*0}$ yield to charged-kaon yield monotonically decreases from peripheral to most-central collisions, as measured by particle multiplicity [$(dN_{ch}/dy)^{1/3}$]. The suppression is observed at the 3$\sigma$ level. This trend can be attributed to re-scattering of the $K^{*0}$ decay daughters during the hadronic phase. The result from this study agrees with previous BES-I observation. The $K^{*0}/K$ ratios in central Au+Au collisions deviate from thermal model calculations, without final-state rescattering with a statistical significance of more than 6$\sigma$, whereas this ratio is consistent with the thermal model for peripheral collisions at $\sqrt{s_{NN}} = 7.7-200$ GeV. The transverse momentum spectra of the $K^{*0}$ meson in central Au+Au collisions, when compared with blast-wave predictions that do not include rescattering effects, show a suppression of the $K^{*0}$ yield at $p_{T} < 1.5$ GeV/$c$, consistent with expectations from the UrQMD model. This observation further supports the hypothesis that re-scattering effects, as opposed to resonance regeneration, is dominant during the hadronic phase of the heavy-ion collision evolution particularly at lower $p_{T}$. The measured $K^{*0}/K$ ratios in Au+Au collisions, together with an appropriate $p+p$ baseline at comparable energies, can provide important constraints on the lifetime of the hadronic phase. Future experiments, such as CBM and sPHENIX, will be valuable in further advancing these studies.

Finally, the $K^{*0}/K$ ratios within a given multiplicity bin show additional suppression at the  BES energies, compared to the highest RHIC and LHC energies. This phenomenon might be caused by differences in the type of hadronic interactions taking place within the medium created in the low energy collisions. While transport model calculations qualitatively support this behavior, additional detailed studies are necessary to clarify the underlying physics mechanisms.



\section*{Acknowledgements}
We thank the RHIC Operations Group and SCDF at BNL, the NERSC Center at LBNL, and the Open Science Grid consortium for providing resources and support.  This work was supported in part by the Office of Nuclear Physics within the U.S. DOE Office of Science, the U.S. National Science Foundation, National Natural Science Foundation of China, Chinese Academy of Science, the Ministry of Science and Technology of China and the Chinese Ministry of Education, NSTC Taipei, the National Research Foundation of Korea, Czech Science Foundation and Ministry of Education, Youth and Sports of the Czech Republic, Hungarian National Research, Development and Innovation Office, New National Excellency Programme of the Hungarian Ministry of Human Capacities, Department of Atomic Energy and Department of Science and Technology of the Government of India, the National Science Centre and WUT ID-UB of Poland, German Bundesministerium f\"ur Bildung, Wissenschaft, Forschung and Technologie (BMBF), Helmholtz Association, Ministry of Education, Culture, Sports, Science, and Technology (MEXT), Japan Society for the Promotion of Science (JSPS), and Agencia Nacional de Investigacion y Desarrollo de Chile (ANID), Chile.





\bibliographystyle{elsarticle-num-names}
\bibliography{sample.bib}













\end{document}